# Point Defects in Two-Dimensional RuCl$_3$

Wenqi Yang[1,‡], Linghan Zhu[2], Yan Lu,[3] Erik Henriksen,[2,4] Li Yang[2, 4*]

[1]Department of Physics, University of Science and Technology of China, Hefei, 230026, China

[2]Department of Physics and Institute of Materials Science and Engineering, Washington University, St. Louis, MO, 63130, USA

[3]Department of Physics, Nanchang University, Nanchang 330031, China

[4]Institute of Materials Science and Engineering, Washington University in St. Louis, St. Louis, MO 63130, USA

**Abstract**

Defects are crucial in determining a variety of material properties especially in low dimensions. In this work, we study point defects in monolayer α-phase Ruthenium (III) chloride (α-RuCl$_3$), a promising candidate to realize quantum spin liquid with nearly degenerate magnetic states. Our first-principles simulations reveal that Cl vacancies, Ru vacancies, and oxygen substitutional defects are the most energetically stable point defects. Besides, these point defects break the magnetic degeneracy: Cl vacancies and oxygen substitutional defects energetically favor the zigzag-antiferromagnetic configuration while Ru vacancies favor the ferromagnetic configuration, shedding light on understanding the observed magnetic structures and further defect engineering of magnetism in monolayer α-RuCl$_3$. We further calculated their electronic structures and optical absorption spectra. The polarization symmetry of optical responses provides a convenient signature to identify the point defect types and long-range magnetic orders.

# I. Introduction

Among emerging two-dimensional (2D) van der Waals (vdW) magnetic materials, α-phase Ruthenium (III) chloride (α-RuCl$_3$) has attracted significant research interest because it potentially hosts the long-pursued Kitaev quantum spin liquid (KQSL) [1-8]. Signatures of the proximate KQSL states and Majorana fermions have been observed in thin-film RuCl$_3$ via polarized terahertz spectroscopy and neutron scattering measurements [9-12]. Besides multilayer structures, substantial efforts have recently been made to achieve monolayer RuCl$_3$ [13-14] since it is expected to be a better platform to instantiate the 2D Kitaev honeycomb model intrinsically. Moreover, theoretical studies predicted nearly degenerate zigzag-antiferromagnetic (ZZ-AFM) and ferromagnetic (FM) orders in monolayer α-RuCl$_3$ [15-17], giving hope for realizing the KQSL model where degenerate magnetic states with low energy barriers are prerequisites [18-20].

On the other hand, defects have been widely observed in fabricated 2D magnetic materials and strongly impact electronic structures and magnetic orders. For example, point and multi-atom vacancies were observed in monolayer CrI$_3$ [21], and they can reduce the local crystal field and enhance the exchange interactions [22-23]. Cr vacancies may enhance the magnetic order and introduce an insulator-to-half metal transition in monolayer CrCl$_3$ [24]. Compared with those 2D structures with "robust" magnetic orders, monolayer RuCl$_3$ is supposed to be more sensitive to defects as its nearly degenerate magnetic orders and potential KQSL state could be easily broken by defects [25]. Therefore, a study on defects and associated material properties of monolayer RuCl$_3$ is crucial for understanding available measurements and searching for the KQSL state in 2D vdW structures.

In this work, we employ first-principles simulations to study point defects and their impacts on the electronic, magnetic, and optical properties of monolayer α-RuCl$_3$. The article is organized in the following way. In section II, we present the computational details and properties of pristine monolayer α-RuCl$_3$. In section III, we introduce the typical point defects and how their formation energies evolve with chemical potential. In section IV, we present the relationship between point defects and magnetic orders. In Section V, the electronic band structures and defect levels are discussed. In Section VI, we show the single-particle optical absorption spectra and the optical signatures of defects. In Section VII, we expand the discussion to oxygen substitutional defects. The results are concluded in Section VIII.

## II. Computational Details and Properties of Pristine Monolayer α-RuCl$_3$

Our calculations are performed by the Vienna Ab initio Simulation Package (VASP) based on density functional theory (DFT). We employ projector-augmented wave (PAW) pseudopotential with the Perdew-Burke-Ernzerhof (PBE) functional to describe the exchanging-correlation interactions [26-27]. We set the plane-wave basis with a cut-off energy of 400 eV with the K-point grid of $15 \times 9 \times 1$ for pristine structures. The vacuum thickness is set to be larger than 18 Å to avoid spurious interaction. We adopt the DFT+U approach to treat the correlations between *d*-orbitals of Ru atoms, and the effective Hubbard parameter is set to be $U_{eff} = U - J = 2$ eV, which was used in previous works [16-17]. Spin-orbit coupling (SOC) is included in all calculations. The atomic and magnetic structures were fully relaxed until the total energy is converged within $10^{-6}$ eV, and the force on each atom is less than 0.005 eVÅ$^{-1}$. To mimic a single-point defect, we adopted a (2×3) supercell containing 24 Ru atoms and 72 Cl atoms. The similarly sized supercell was employed in studying point defects in other 2D magnetic materials, such as CrI$_3$ and CrCl$_3$, [28-29] and the converged results of larger 3×3 supercells are presented in Section IV of the supplementary information [30]. Correspondingly, a smaller K-point grid of $3 \times 3 \times 1$ is used for the total-energy calculations of supercells. The K-point grid for the optical absorption spectra is $9 \times 9 \times 1$ for the supercell.

We start from four types of magnetic configurations: ZZ-AFM, FM, Néel-antiferromagnetic (Néel-AFM), and stripy-antiferromagnetic (Stripy-AFM). [31] The top views of these magnetic configurations are summarized in Figures 1 (a)-(d). The total-energy calculation confirms that the ZZ-AFM configuration is the ground state. Their total energies follow the order of $E_{ZZ\text{-}AFM} < E_{FM} < E_{stripy} < E_{Néel}$. This agrees with previous publications. [16-17] It is important to note that the total energy of the FM order is only 0.034 meV/f.u. higher than that of the ZZ-AFM one, which is within the accuracy of the DFT simulation. Such nearly degenerate ground states [32] support the idea that monolayer α-RuCl$_3$ may be in a proximate KQSL phase. The detailed relaxed structure parameters are listed in Table S1 of Section I of the supplementary information [30].

### III. Point Defects and Formation Energies

Beyond the pristine structure, we consider six types of point defects, which have been typically observed/studied in other 2D magnetic materials, e.g., CrI$_3$ and CrCl$_3$ [23][28]. These defects can be classified into three categories: 1) vacancy defects: the single Cl atom vacancy ($V_{Cl}$) or single Ru atom vacancy ($V_{Ru}$), which are plotted in Figures 2 (a) and (b), respectively; 2) antisite defects: a Ru atom takes the place of a Cl atom ($Ru_{Cl}$) or a Cl atom takes the place of a Ru atom ($Cl_{Ru}$), which are plotted in Figures 2 (c) and (d), respectively; 3) adatom defects: an excess Cl atom locates on top of the material ($Cl_{on}$) or an excess Ru atom locates on top of the material ($Ru_{on}$), which are plotted in Figures 2 (e) and (f), respectively.

The thermal stability of these point defects can be obtained by calculating the formation energy ($E_{form}$) based on the standard approach of Ref. [33]:

$$E_{form}[X] = E_{tot}[X] - E_{tot}[bulk] - \sum_i n_i \mu_i,$$

where $E_{tot}[X]$ is the total energy of the supercell containing the type-$X$ defect, and $E_{tot}[bulk]$ is the total energy of pristine monolayer α-RuCl$_3$ without defects. In addition, the integer index $n_i$ denotes the change in the number of type-$i$ atoms. $\mu_i$ represents the chemical potential of a single type-$i$ atom. To prevent Ru and Cl atoms from forming the bulk Ru precipitate or gaseous $Cl_2$ phases during the process of material growth, the chemical potential of Ru atoms should not exceed that of the bulk Ru, and the chemical potential of Cl atoms cannot exceed that of the gaseous $Cl_2$. [28] Besides, the chemical potentials of Ru, Cl, and $RuCl_3$ follow the relationship $\mu_{Ru} + 3\mu_{Cl} = \mu_{RuCl_3}$. Then, the range of $\mu_{Cl}$ values is set as $-0.96$ eV $\leq \mu_{Cl} \leq 0$ eV. By inserting this range into the formula of formation energy, we can obtain the defect formation energy, $E_{form}$, as a function of the chemical potential of Cl, *i.e.*, $\mu_{Cl}$.

The defect formation energies vs the Cl chemical potential are summarized in Figure 3 (a). Due to the low mobility of our studied defects (See Section II of the supplementary information [30]), the contributions of kinetic effects to the thermodynamic stability of defects can be ignored. Ru and Cl point vacancies have the lowest formation energies, so they are the most stable defect types, and the Cl antisite ($Cl_{Ru}$) and adatom ($Cl_{on}$) defects are generally the second most stable. Other types of defects are the least stable ones. The relative stability of defects is also sensitive to the chemical potential. The Cl-rich condition favors the formation of Ru vacancies ($V_{Ru}$) while the Cl-poor condition favors the formation of Cl vacancies ($V_{Cl}$). The critical chemical potential of Cl is found to be -0.79 eV between these two types of vacancies.

If we consider a charge state q, the expression of formation energy is changed into:

$$E_{form}[X^q] = E_{tot}[X^q] - E_{tot}[bulk] - \sum_i n_i \mu_i + q\epsilon_F + E_{corr},$$

where $\epsilon_F$ is the Fermi level, and $E_{corr}$ is the correction term which accounts for the finite-size supercell correction. [34-37] We consider two main parts of the correction term: Firstly, there are electrostatic interactions between the defect and its periodic images, which can be estimated from the Madelung energy of an array of point charges with neutralizing background [34][35][38]:

$$E^{Md} = -\frac{\alpha q^2}{2\varepsilon L},$$

where $\alpha$ is the Madelung constant, and L is the supercell lattice constant. Secondly, we need to align the electrostatic potentials of the intrinsic structure and the defective structure, the contribution of which to the correction term is [38][39]:

$$E^{alignment} = -q\Delta V_{q/b},$$

where $\Delta V_{q/b}$ can be estimated through the comparison of the electrostatic potential in the region far from the charged defect and in the intrinsic structure calculation: $\Delta V_{q/b} = V_{q|far} - V_b$.

The formation energies of Cl vacancy ($V_{Cl}$) and Ru vacancy ($V_{Ru}$) at different charge states are presented in Figures 3 (c) and (d). As the Fermi level varies within the range of the band gap, [33] the charge-neutral defects keep the most stable ones. Moreover, previous works show that the doping levels of defects in 2D magnetic insulators can be very deep. [40] Combined with experimental observations that most fabricated $RuCl_3$ is strongly insulating, we will focus on the charge-neutral defects in this work.

### IV. Magnetic Structures of Point Defect States

As shown in Figures 1 (a) and (b), the ZZ-AFM and FM orders are nearly degenerate in α-$RuCl_3$. However, defects may break this degeneracy and change the magnetic ground order. Motivated by this idea, we have calculated the total energy of monolayer α-$RuCl_3$ with point defects as listed in Figure 2. The magnetic ground state of defect structures is indeed changed. Figure 3 (b) shows the relative energy difference between the FM and ZZ-AFM states in a (2×3) supercell for different defects. Cl vacancies ($V_{Cl}$) prefer the ZZ-AFM phase while Ru vacancies ($V_{Ru}$) prefer the FM phase. For those antisite cases, $Ru_{Cl}$ defects favor the ZZ-AFM state while $Cl_{Ru}$ defects favor the FM configuration. For adatom defects, the system always favors the ZZ-AFM configuration.

The above energy differences depend on the simulated defect density. One defect in a 2×3 supercell corresponds to a high defect density (~$2.5 \times 10^{13}$ cm$^{-2}$). To date, there is no quantitative experimental result of the point defect density in α-RuCl$_3$. The typical defect density is around $10^{12}$ cm$^{-2}$ in CVD-grown transition metal dichalcogenides (TMDs) [41-42]. Because our studied defect structures are insulators (discussed in section V), the spin exchange coupling is the dominant factor in determining the magnetic ground states. In Section III of the supplementary information [30], we present the values of the spin exchange couplings. The strength of the exchange coupling decreases rapidly as the distance increases (the decay length is 1.67 Å). If the distance is greater than three Ru-Ru bond lengths, the coupling strength is about two orders of magnitude smaller than the nearest-neighboring coupling. This indicates that the spin exchange coupling is short-range, which is consistent with the results obtained using larger 3×3 supercells (as discussed in Section IV of the supplementary information [30]). As a result, high-density (~$10^{13}$ cm$^{-2}$) defects can decide the global magnetic order. For low-density (less than $10^{12}$ cm$^{-2}$) defects, the average distance between defects is more than 10 nm, which is much larger than the decay length of spin exchange coupling. In this situation, the overall magnetic order will revert to the intrinsic ones while point defects will merely exhibit local magnetic configurations around defects. Finally, as a quenched disorder, defects may hinder the observation of QSL states. [18][25] Thus, experimental efforts shall try to purify RuCl$_3$ to search for QSL.

We also study magnetic properties near the vacancies. As stated before, the magnetic moments are along the out-of-plane direction in the pristine structure. Point vacancies can change the magnetic moment and induce in-plane magnetic moments on Ru atoms around vacancies. The detailed magnetic moment distributions are depicted in section V of the supplementary information [30].

Finally, as shown in Figure 3 (a), the defect type and density can be tuned by the chemical potential during growth. Therefore, it is possible to control magnetic orders via defect engineering. For example, Cl-poor conditions (Cl chemical potential less than 0.79eV), can help to form Cl point vacancies and enhance the ZZ-AFM configuration. Cl-rich conditions (Cl chemical potential larger than 0.79eV) can support the formation of Ru point vacancies and may prefer the FM ordering.

## V. Electronic Structures of Point Defect States

We begin with the electronic band structures of pristine monolayer ZZ-AFM α-RuCl$_3$, which are presented in Figure 4 (a). The DFT+U band gap is around 1 eV. The band edges are very flat due to the significant *d*-orbital component from Ru atoms, as shown in the projected density of states (PDOS). We also notice that the absolute energies of the conduction band minimum (CBM) and valence band maximum (VBM) are significantly lower than many other 2D materials. For example, the DFT-calculated CBM of most TMDS is around -3 to -4 eV, [43-44], and the Fermi level of graphene is around -4.6 eV. However, the CBM of monolayer α-RuCl$_3$ is around -5.5 eV. This deep work function is in good agreement with the observations of strong charge transfer and electron doping in thin-film RuCl$_3$/graphene heterostructures. [15, 45-49]

Next, we work on the electronic structures with point defects included. To better illustrate the impact of point defects, we have utilized the band-unfolding method to convert the band structures of supercells containing point defects into the first Brillouin zone of the conventional unit cell [50]. This method not only facilitates the identification of impurity energy levels but also serves as a reference for angle-resolved photoemission spectroscopy measurements. By comparing the unfolded band structures with the pristine band structures, we can determine the impurity energy levels, whose intensity is proportional to the defect density in the samples. Since point vacancies are the most stable defects as shown in Figure 3 (a), the following discussion will be confined to the Ru and Cl point vacancies. For all defects, the system stays insulating. Figure 4 (b) presents the result of Ru vacancies. Many unoccupied defect levels emerge close to the CBM, and they are mainly from the Ru orbitals. The band gap is reduced from the intrinsic value of 1 eV to around 0.62 eV.

The situation of Cl vacancies is more complicated. The reason is that, for the ZZ-AFM order, there are two subtly different positions for Cl vacancies. One is in the zigzag chain and is named the type A defect. The other is between zigzag chains and is named the type B defect. Their atomic structures and the AFM zigzag chains are shown in Figures 4 (e) and (f), respectively. The type A Cl vacancy has slightly lower formation energy (~ 40 meV) and is more energetically favored. However, their energy difference is small, and both types of Cl vacancies will exist in fabricated samples.

As shown in Figure 4 (c), many in-gap defect levels emerge, and the band gap is reduced to 0.64 eV for type-A Cl vacancy. Unlike those of Ru vacancies, these defect states are fully occupied and are close to the VBM. On the other hand, for the type-B Cl vacancy, the electronic structure in Figure 4 (d) is qualitatively different. There is mainly one in-gap occupied defect level. The band gap is about 0.48 eV, which is smaller than that of the Type-A Cl vacancy defect. This indicates that the type-B Cl vacancies may be more visible in photoluminescence (PL) measurements because of their smaller band gap. This provides a unique approach to identify the defect types in experiment.

We then turn to the FM order. The electronic structure of the pristine structure is presented in Figure 5 (a). The band gap is around 0.75eV, which is sizably smaller than that of the ZZ-AFM order. In addition, the conduction band edge of the FM phase is more dispersive than that of the ZZ-AFM state; the electron effective mass of the FM phase is around 1.99 $m_e$ while that of the ZZ-AFM phase is around 10.92 $m_e$. Finally, because of the small energy difference between FM and ZZ-AFM phases, an external magnetic field can switch the ZZ-AFM to the FM phase and further tune the transport and optical properties. These fundamental electronic quantities of the pristine structure are summarized in Table S2 of Section I of the supplementary information [30], and the effect of SOC on band structures is shown in Section VI of the supplementary information [30].

The electronic structures for FM defect states are shown in Figures 5 (b) and (c) for Ru and Cl vacancies, respectively. Generally, we observe similar defect levels as those in the ZZ-AFM order, and the structures maintain insulating. For example, Ru vacancies generate defect levels close to CBM, and Cl vacancies generate defect levels close to VBM. The band gap is reduced to 0.65 eV and 0.5 eV for the Ru and Cl vacancies, respectively.

It is worth mentioning that, although DFT is known for underestimating the band gaps of materials, the above-discussed relative energy differences between different magnetic orders are reliable. For example, the advanced GW calculation [51-53] shows that, despite the different band structures, the many-electron self-energy corrections are similar for the defect levels and bulk states in monolayer TMDs, [54] and they are not sensitive to the FM or AFM orders in monolayer and multiple-layer $CrCl_3$. [55]

**VI. Optical Absorption Spectra**

Point defects are important in deciding optical responses of 2D materials, such as TMDs. [56-59] In this section, we present the single-particle optical absorption spectra to provide signatures for experimental identifications of defects in 2D α-RuCl$_3$. Similarly, we focus on the vacancy defects with the ZZ-AFM and FM orders because of their energetical stability.

Figure 6 (a) shows the optical absorption spectra of pristine ZZ-AFM monolayer RuCl$_3$. The absorption edge is around 1 eV, which is consistent with the band gap. Importantly, the spectrum is highly anisotropic; the absorption along the zigzag direction is stronger than that along the armchair direction. These similar anisotropic optical spectra were also observed in 2D ZZ-AFM FePS$_3$ [60]. After including defects, new absorption peaks appear because of those in-gap defect states. For the Ru vacancies shown in Figure 6 (b), a new peak shows around 0.7 eV, which is about 300 meV below the bulk band gap. Its dipole oscillator strength is partially anisotropic. As shown in the inset, the dipole oscillator strength of the peak at 0.7 eV is stronger along the zigzag chain direction, and the ratio between the minimum and maximum is about 0.4.

The different electronic structures of the two AFM Cl vacancies as shown in Figures 4 (e) and (f) induce distinct optical spectra. In Figure 6 (c), for the type-A Cl vacancy, the lowest-energy defect peak is around 0.66 eV. However, for the type-B Cl vacancy, the lowest energy defect peak is substantially lower (~ 0.48 eV) in Figure 6 (d). More importantly, the polarization is sensitive to the location of Cl vacancies. In Figure 6 (c), the polarization of the type-A Cl vacancy is roughly along the zigzag chain but deviated about 30 degrees. We notice that there are four degenerate positions for type-A Cl vacancies. The overall polarization will be their superpositions, and the averaged polarization of type-A defects will be along the zigzag chains. (See Figure S5 of the supplementary information [30]). In Figure 6 (d), the polarization of the type-B Cl vacancy is perpendicular to the zigzag chain. Particularly, the lowest-energy peak of the type-B Cl vacancy is about 180 meV lower than that of the type-A Cl vacancy. Therefore, the observed PL will be dominated by type-B defects and exhibit significant anisotropy. This is useful to identify the defect type, the zigzag AFM order, and its direction.

Then we turn to the FM-order RuCl$_3$. The optical absorption spectrum of the pristine structure is presented in Figure 7 (a). The absorption edge starts from around 0.75 eV, which agrees with the band gap. Unlike the ZZ-AFM case, the optical absorption spectrum is isotropic concerning the polarization direction. The optical absorption spectra for Ru and Cl defects are presented in Figures 7 (b) and (c), respectively. For Ru vacancies, a new absorption peak emerges from 0.65 eV, which is from the transition between VBM to the lowest unoccupied defect level. Importantly, its dipole oscillator strength is isotropic because of the C$_3$ symmetry. For Cl vacancies, the lowest peak is at 0.59 eV, which is from the transition between the highest occupied defect level to CBM. Interestingly, it is highly anisotropic, and the preferred polarization direction is roughly along the broken Ru-Cl-Ru bond. It is worth mentioning that the anisotropic spectrum in Figure 7 (c) may not be observed in experiments. Because of the C$_3$ symmetry of monolayer α-RuCl$_3$, there are six degenerate bond-breaking directions (See Figure S6 of supplementary information [30]). In real samples, the combination of their contributions will result in an isotropic signal.

We address that the DFT-calculated optical spectra in Figures 6 and 7 are based on single-particle inter-band transitions. It is known that many-electron effects, such as electron-hole (*e-h*) interactions or excitonic effects, are significant in 2D materials. [61-62] Therefore, the absolute energy and profile of the optical spectra in Figures 6 and 7 will not align with experimental observations. On the other hand, as shown in previous first-principles simulations, the exciton-induced energy shifts for defect and bulk transitions are not significantly different. For example, the *e-h* binding energy of the defect level from Se vacancies in monolayer MoSe$_2$ is about 0.6 eV which is nearly the same as that of the pristine structure. [63] Therefore, we expect the relative energies of defect peaks in Figures 6 and 7 to be qualitatively correct. More importantly, *e-h* interactions will not change polarization and symmetry. For example, the valley polarization dependency of TMD and the anisotropic spectrum of black phosphorus are the same for single-particle and two-particle results. [64-67] In this sense, DFT-predicted polarization properties provide a convenient way of identifying defect types. Moreover, because the local field effect is not included in DFT simulations, we only present the optical absorption spectra for in-plane polarized light to avoid the depolarization effect.

Finally, scanning tunneling microscopy (STM) is another powerful tool for studying defects in 2D materials. We have simulated STM images of Cl vacancies and Ru vacancies under a -0.7 V bias, which is widely adopted for STM simulations [28][68]. As depicted in the upper-right corner inserts of Figures 6 and 7, the main STM feature is determined by the Ru atoms (lattice). The Ru vacancies are easily seen as missing bright spots, while the Cl vacancies are not as significant. The vacancies will also affect the contrast of nearby Ru atoms. As shown in Figures 6(b) and 7(b), the images of nearby Ru atoms are dimmer than others, which is more significant in the ZZ-AFM phase than in the FM phase. A similar effect is also observed for the Cl vacancies in the FM phase (Figure 7(c)). However, in the ZZ-AFM phase, the Cl vacancies make the images of nearby Ru atoms brighter than others (Figures 6(c) and (d)).

**VII. Oxygen Point Defects**

We expand our point defect discussion to oxygen impurities, which are important in 2D materials because they can form complicated defect structures and result in degradation. [69] Since the compound $RuOCl_2$ is known to exist, [70] we focus on the oxygen substitutional point defect, $O_{Cl}$, where an O atom takes the place of a Cl atom. The atomic structure of $O_{Cl}$ is presented in Figure 8 (a). Based on DFT calculations, we find the system highly prefers oxygen adsorption on Cl vacancies because the energy gain is about 2.7 eV.

We have calculated the magnetic order, electronic structure, and optical absorption of the $O_{Cl}$ vacancy. We find that, after the Cl vacancy adsorbs an O atom, the system still prefers the ZZ-AFM order, and the energy difference between ZZ-AFM and FM is increased to be about 0.14 meV/f.u. for a 2×3 supercell (~ density of $2.5 \times 10^{13}$ cm$^{-2}$). This energy gain is smaller than that of a single Cl vacancy (~ 0.38 meV /f.u.) but larger than that of the pristine structure (~ 0.03 meV/f.u.), indicating that oxygen substitutional defects enhance the ZZ-AFM ground state. Since the O atom can absorb two types of Cl vacancies in the ZZ-AFM structure, we correspondingly denote the $O_{Cl}$ defects as type A and type B. The band structures and PDOS are presented in Figures 8 (b)-(d) for the FM and two types of ZZ-AFM orders. Generally, we observe in-gap defect levels which are fully occupied and are close to VBM. Different from Cl vacancies, there are fewer defect levels in the case of $O_{Cl}$. Moreover, PDOS shows that most of the impurity states are from Ru atoms, and the contributions of Cl atoms and the O atom are minor.

The optical absorption spectra are substantially different from those of Cl vacancies. We first focus on the FM phase. As shown in Figure 8 (e), the optical absorption spectrum is similar to the intrinsic one with slight anisotropy. This indicates that the optical transition between the in-gap occupied defect level and CBM is relatively dark. In the ZZ-AFM phase, for the type-A defect, Figure 8 (f) shows that the optical spectrum is nearly the same as that of the pristine structure. There is no visible defect peak because the optical transition from the occupied defect level to CBM is dark. For the type B defect, an isotropic defect peak is observed in Figure 8 (g). It is strongest along the perpendicular direction of the zigzag chain although the intensity of this side peak is weaker than that of Cl vacancies. Because of the lower energy of the defect peak in Fig. 8 (g), the PL measurement of $O_{Cl}$ structures will be dominated by the type-B defects and exhibit the anisotropic character.

## VIII. Conclusion and Summary

In this study, we investigated point defects in 2D α-RuCl$_3$, a promising material for realizing Kitaev quantum spin liquids. Using first-principles simulations, we found that Cl and Ru vacancies are the most energetically stable point defects, which can influence the material's magnetic order. This sheds light on the complex magnetic phases observed in α-RuCl$_3$ and demonstrates the potential of defect engineering for controlling magnetism. Furthermore, we calculated the electronic structures and optical absorption spectra of defect states. The symmetry of the optical responses associated with defects can be used to distinguish different point defect types and magnetic orders. Finally, we expanded our investigation to include oxygen point defects, which were found to be highly energetically favorable and mainly contribute to optically dark states.

**Acknowledgment**

We thank Kenneth S. Burch for fruitful discussions. W.Y. acknowledges supercomputer support from the Supercomputing Center of University of Science and Technology of China. L.Z. is supported by the National Science Foundation (NSF) grant No. DMR-2124934. Y.L. is supported by the National Natural Science Foundation of China grant No. 12164026. E.A.H. and L.Y. are supported by the Office of the Under Secretary of Defense for Research and Engineering under award number FA9550-22-1-0340. E.A.H. acknowledges support from the Moore Foundation Experimental Physics Investigators Initiative award no. 11560. L.Y. acknowledges Extreme Science and Engineering Discovery Environment (XSEDE), which is supported by NSF grant number ACI-1548562. The authors acknowledge the Texas Advanced Computing Center (TACC) at The University of Texas at Austin for providing HPC resources.

**Figures:**

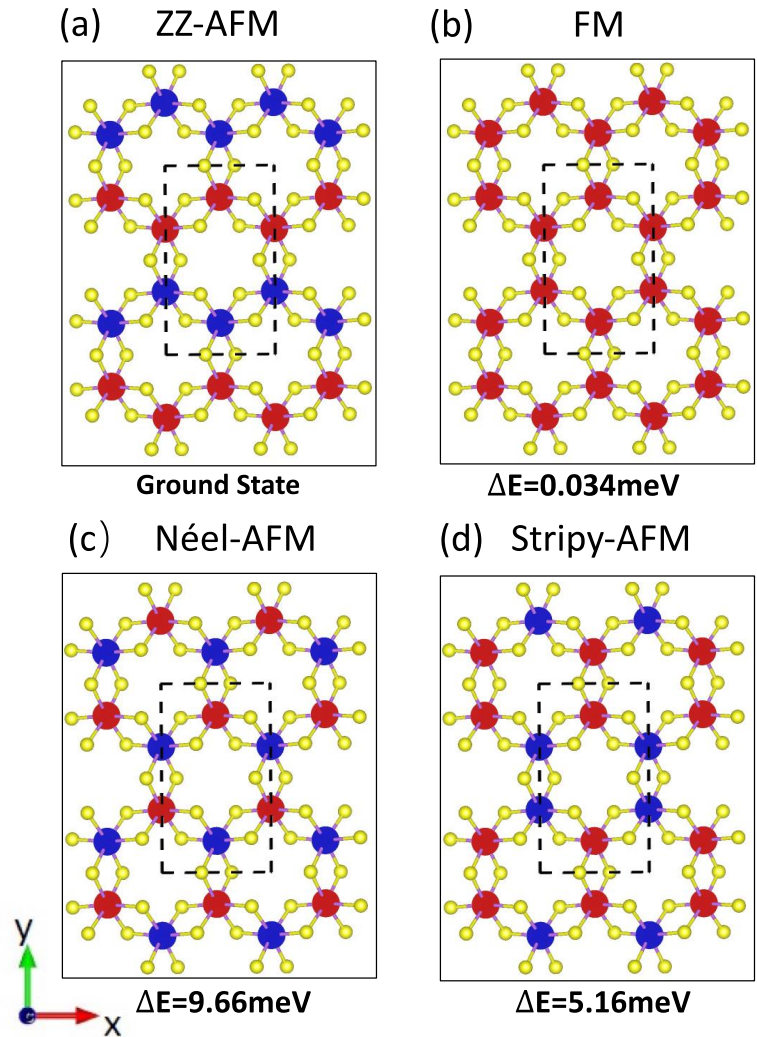

**Figure 1** Top view of magnetic configurations of monolayer α-RuCl$_3$: (a) ZZ-AFM, (b) FM, (c) Néel-AFM, and (d) Stripy-AFM. The blue and red colors represent the spin-up and spin-down Ru atoms, respectively. The Cl atoms are marked by the yellow color. The unit cell is marked by the dashed square, which is twice as large than the primitive cell because we need to include those AFM orders. The total energy difference per formula cell is marked at the bottom of each figure, and the zero-reference point is based on the ZZ-AFM.

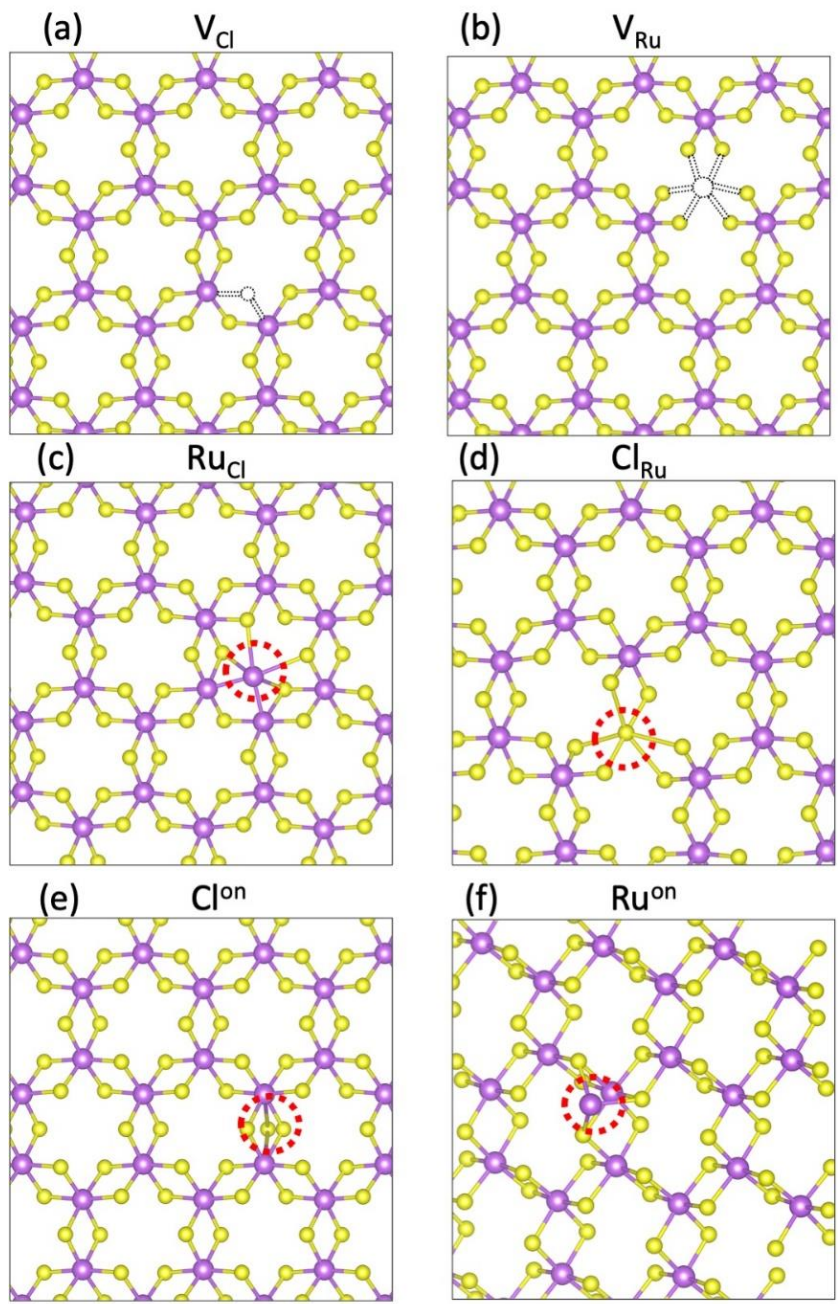

**Figure 2** Top view of the atomic structure of point defect. (a) Single Cl atom vacancy ($V_{Cl}$), (b) single Ru atom vacancy ($V_{Ru}$), (c) a Ru atom takes the place of a Cl atom ($Ru_{Cl}$), (d) a Cl atom takes the place of a Ru atom ($Cl_{Ru}$), (e) an excess Cl atom locates on the top of the layer ($Cl^{on}$), (f) an excess Ru atom locates on the top of the layer ($Ru^{on}$). The missing atom is marked by black dashed lines. The red dashed circle is used to guide readers' eyes to the point defect.

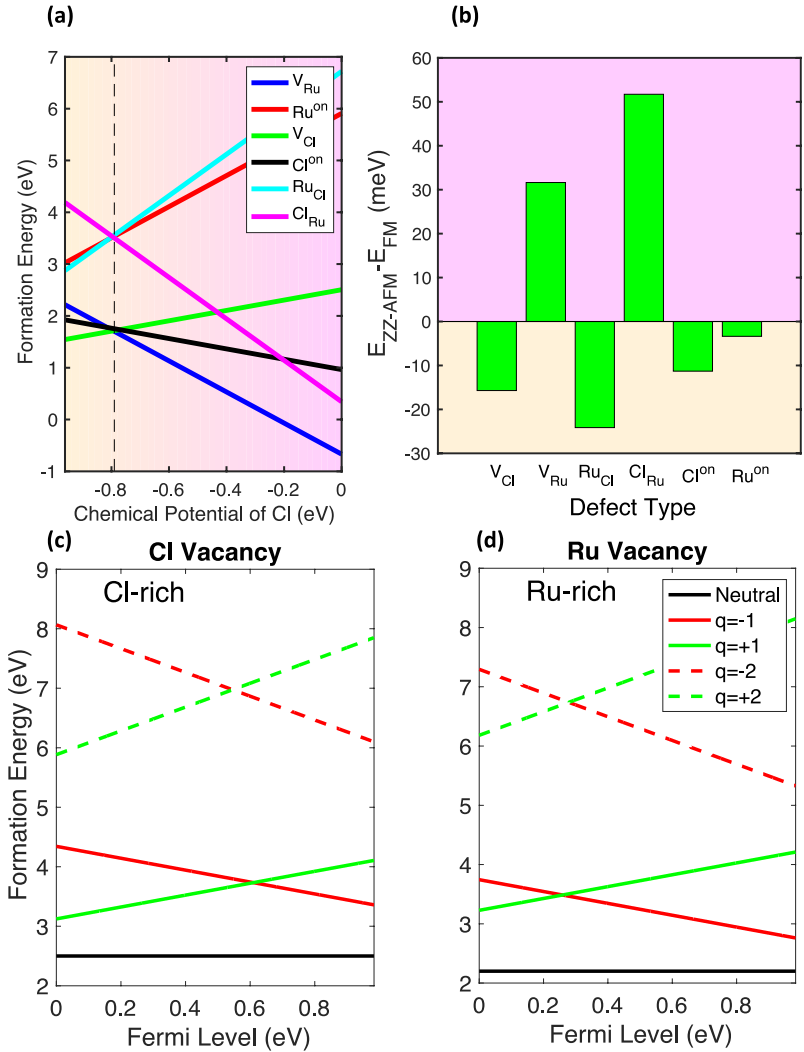

**Figure 3** (a) Relationships between defect formation energies and the Cl chemical potential. The lines in different colors represent different types of point defects. (b) Energy splitting between FM phase and ZZ-AFM phase induced by defects (2x3 supercell). Each bar represents a defect type. Background color represents the magnetic phase: pink means the system favors the FM order, and yellow means the system favors the ZZ-AFM order. Formation energies of defects at different charge state as a function of the Fermi level: (c) formation energy of the Cl vacancy, (d) formation energy of the Ru vacancy.

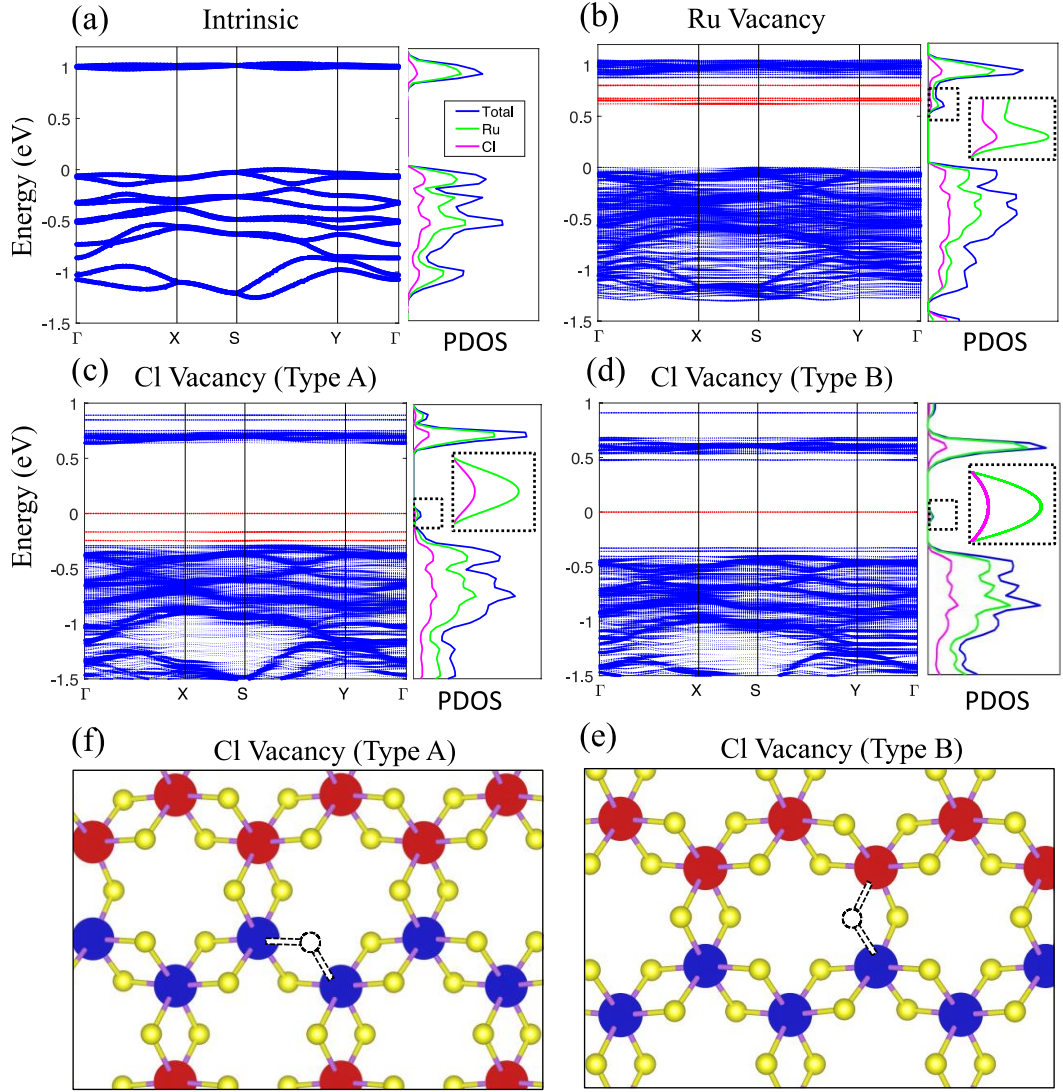

**Figure 4** Unfolded-Band structures and PDOS of the ZZ-AFM phase: (a) Pristine structure, (b) Ru vacancy ($V_{Ru}$), (c) Type A Cl vacancy, (d) Type B Cl vacancy. We use red color to highlight the impurity bands in band structures. The green lines and pink lines represent the PDOS of Ru atoms and Cl atoms, respectively. The PDOS of impurity states is circled with dotted rectangles and magnified. The zero energy is set to be the highest occupied state. Figures (e) and (f) show top views of the atomic structures of the type A defect and the type B defect, respectively. The blue and red colors represent the spin-up and spin-down Ru atoms.

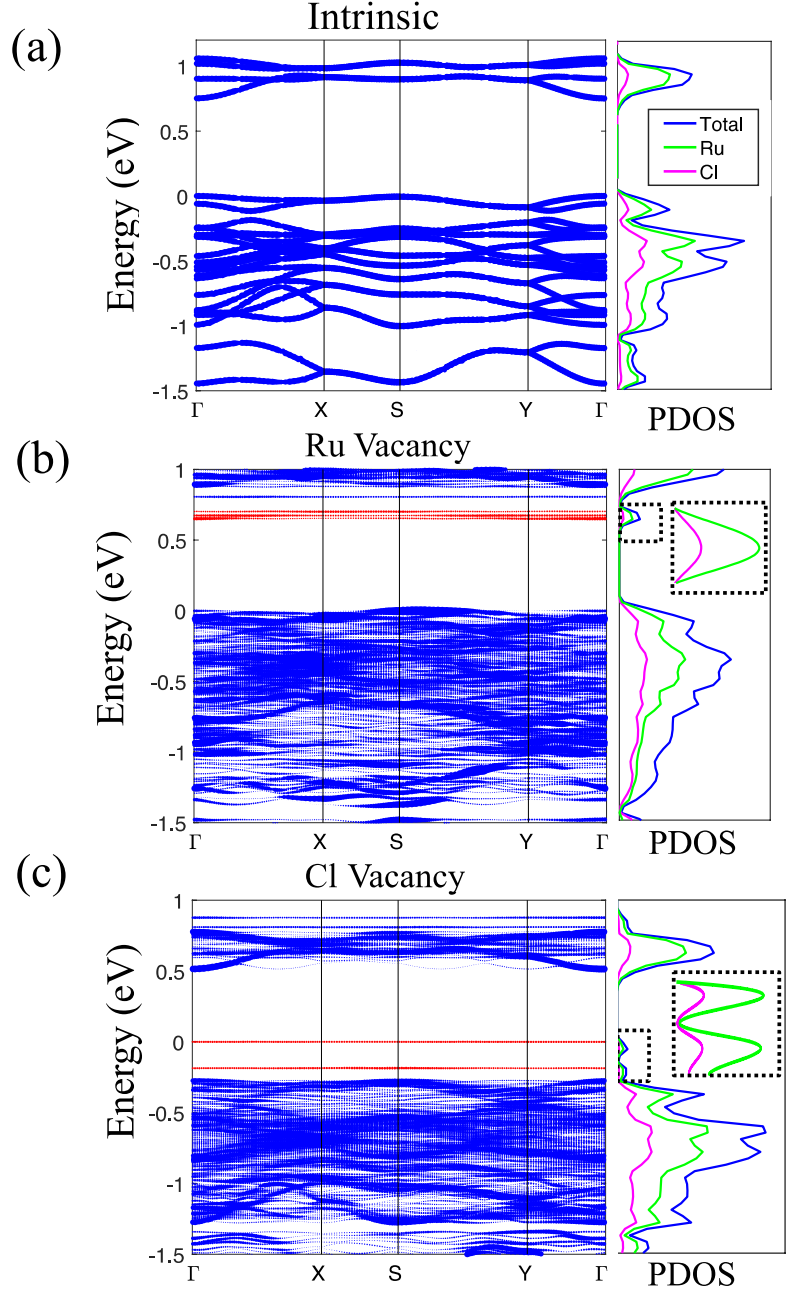

**Figure 5** Unfolded-Band structures and PDOS of the FM phase. (a) Pristine structure, (b) Ru vacancy ($V_{Ru}$), and (c) Cl vacancy ($V_{Cl}$). The red color is used to highlight the impurity bands in band structures. The green lines and pink lines represent the PDOS of Ru atoms and Cl atoms, respectively. The PDOS of impurity states is circled with dotted rectangles and magnified. The zero energy is set by the highest occupied state.

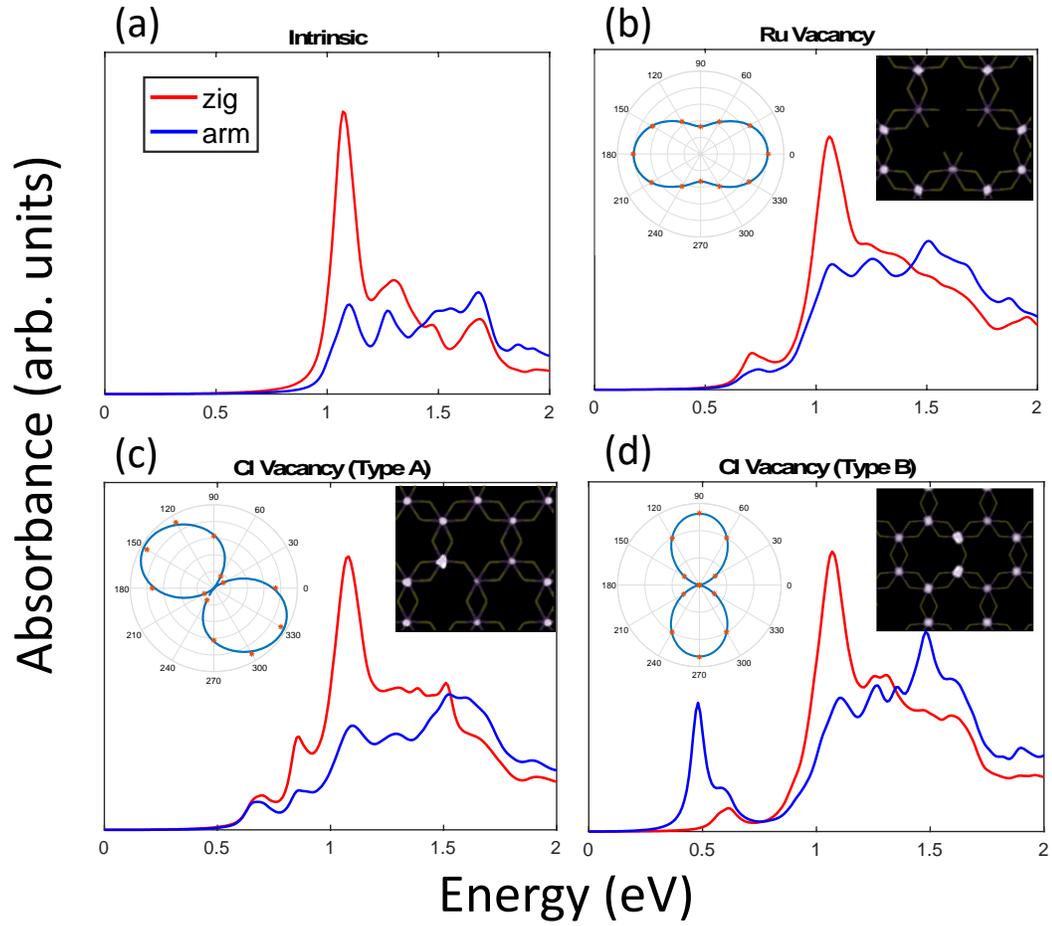

**Figure 6** Optical absorption spectra of ZZ-AFM RuCl$_3$: (a) Pristine structure, (b) Ru vacancy (V$_{Ru}$), (c) type A Cl vacancy, and (d) type B Cl vacancy. Red lines and blue lines represent the polarization of incident light along the zigzag direction and armchair direction, respectively. The polarization dependence of the lowest-energy peak is plotted in the upper-left corner insert. The simulated STM images of defect structures are shown in the upper-right corner inserts. 50-meV smearing is applied in these spectra.

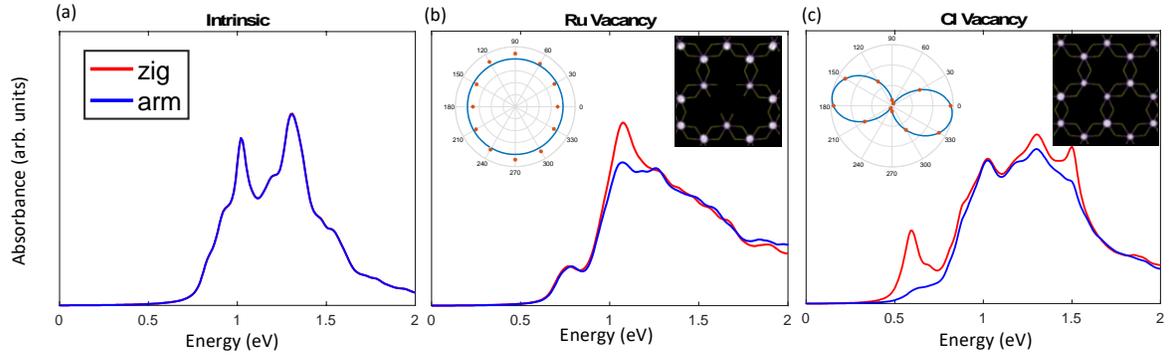

**Figure 7** Optical absorption spectra of FM RuCl$_3$: (a) pristine structure, (b) Ru vacancy (V$_{Ru}$), and (c) Cl vacancy (V$_{Cl}$). Red lines and blue lines represent the polarization of incident light along the zigzag direction and armchair direction, respectively. The polarization dependence of the lowest-energy peak is plotted in the upper-left insert. The simulated STM images of defect structures are shown in the upper-right corner inserts. 50-meV smearing is applied in these spectra.

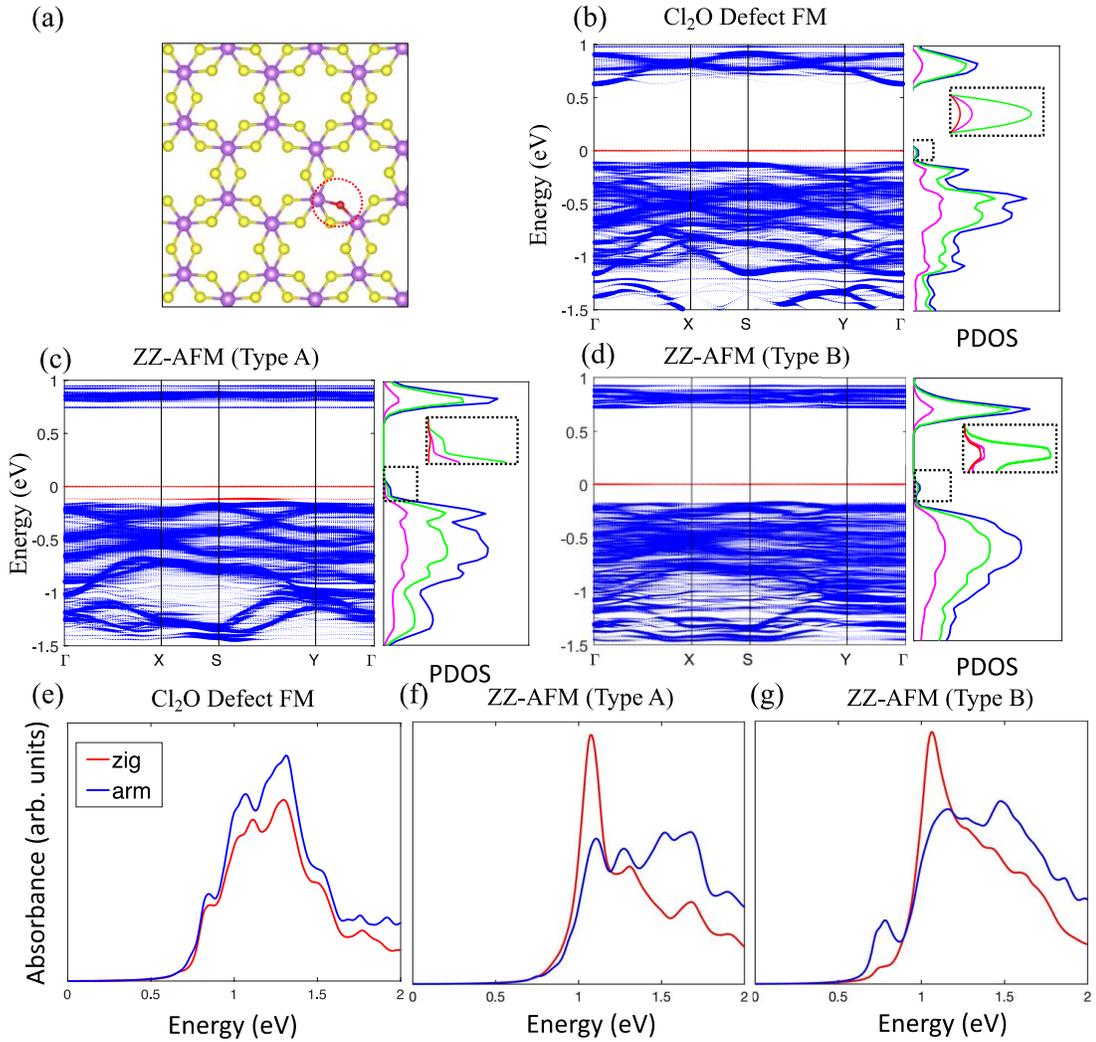

**Figure 8** (a) Top view of $O_{Cl}$ defect atomic structure, in which an O atom takes the place of a Cl atom. (b) Unfolded-Band structures and PDOS of FM $RuCl_3$ with an $O_{Cl}$ defect. (c) and (d) those of type A and type B $O_{Cl}$ defects in the ZZ-AFM phase. We use the red color to highlight the impurity bands in band structures. The green lines and pink lines represent the PDOS of Ru atoms and Cl atoms, respectively. The PDOS of impurity states is circled with dotted rectangles and magnified in the insert, where the red lines represent the PDOS of the absorbed O atom. The zero energy is set to be the highest occupied state. (e) Optical absorption spectra of FM $RuCl_3$ with $O_{Cl}$ defects. (f) and (g) Those of type A and Type B defects in ZZ-AFM $RuCl_3$. 50-meV smearing is applied in these spectra.

[30] See Supplemental Material at [URL] for structure and electronic parameters of $\alpha - RuCl_3$; energy barrier for the Cl vacancy; spin exchange coupling constants; supercell size effects; magnetic moments around point defects; spin-orbit coupling in the band structure of FM monolayer $\alpha - RuCl_3$; overall optical polarization of Cl vacancies.

[31] S. Sarikurt, Y. Kadioglu, F. Ersan, E. Vatansever, O. Üzengi Aktürk, Y. Yüksel, Ü. Akinci, and E. Akürk, Chem. Phys. 20, 997 (2018).

[32] H. Yang, B. Gong, K. Liu, and Z. Lu, J. Phys.: Condens. Matter 31, 025803 (2019).

[33] S. B. Zhang and J. E. Northrup, Phys. Rev. Letter 10, 2339 (1991).

[34] H. P. Komsa, T. T. Rantala, and A. Pasquarello, Phys. Rev. B 86, 045112 (2012).

[35] T. R. Durrant, S. T. Murphy, M. B. Watkins, and A. L. Shluger, J. Chem. Phys. 149, 024103 (2018).

[36] T. J. Smart, F. Wu, M. Govoni, and Y. Ping, Phys. Rev. Materials 2, 124002 (2018).

[37] F. Wu, G. Andrew, R. Sundararaman, D. Rocca, and Y. Ping, Phys. Rev. Materials 1, 071001 (2017).

[38] C. Freysoldt, J. Neugebauer, and C. G. Van de Walle, Phys. Rev. Lett. 102, 016402 (2009).

[39] R. Sundararaman, and Y. Ping, J. Chem. Phys. 146, 104109 (2017).

[40] D. Wang, D. Han, X. B. Li, S. Y. Xie, N. K. Chen, W. Q. Tian, D. West, H.-B. Sun, and S. Zhang, Phys. Rev. Lett. 114, 196801 (2015).

[41] M. R. Rosenberger, H. J. Chuang, K. M. McCreary, C. H. Li, and B. T. Jonker, ACS Nano. 12, 1793 (2018).
26